\documentclass[]{article}

\usepackage{graphicx}
\usepackage{subcaption}
\usepackage{mwe}

\usepackage{lmodern,float,xcolor}
\usepackage{amssymb,amsmath}
\usepackage{ifxetex,ifluatex}
\usepackage{fixltx2e} 
\ifnum 0\ifxetex 1\fi\ifluatex 1\fi=0 
  \usepackage[T1]{fontenc}
  \usepackage[utf8]{inputenc}
\else 
  \ifxetex
    \usepackage{mathspec}
  \else
    \usepackage{fontspec}
  \fi
  \defaultfontfeatures{Ligatures=TeX,Scale=MatchLowercase}
\fi
\IfFileExists{upquote.sty}{\usepackage{upquote}}{}
\IfFileExists{microtype.sty}{%
\usepackage{microtype}
\UseMicrotypeSet[protrusion]{basicmath} 
}{}
\usepackage[margin=1in]{geometry}
\usepackage{hyperref}
\hypersetup{unicode=true,
            pdftitle={Draft: Assessing the Impact of Gamification on Self-Directed Learning in Medical Students},
            pdfauthor={Lee De Zhang, Vik Gopal, Chan Jia Min, Ng Li Shia and Ang Eng Tat},
            pdfborder={0 0 0},
            breaklinks=true}
\usepackage{graphicx,grffile}
\makeatletter
\def\maxwidth{\ifdim\Gin@nat@width>\linewidth\linewidth\else\Gin@nat@width\fi}
\def\maxheight{\ifdim\Gin@nat@height>\textheight\textheight\else\Gin@nat@height\fi}
\makeatother
\setkeys{Gin}{width=\maxwidth,height=\maxheight,keepaspectratio}
\IfFileExists{parskip.sty}{%
\usepackage{parskip}
}{
\setlength{\parindent}{0pt}
\setlength{\parskip}{6pt plus 2pt minus 1pt}
}
\providecommand{\tightlist}{%
  \setlength{\itemsep}{0pt}\setlength{\parskip}{0pt}}
\ifx\paragraph\undefined\else
\let\oldparagraph\paragraph
\renewcommand{\paragraph}[1]{\oldparagraph{#1}\mbox{}}
\fi
\ifx\subparagraph\undefined\else
\let\oldsubparagraph\subparagraph
\renewcommand{\subparagraph}[1]{\oldsubparagraph{#1}\mbox{}}
\fi

\let\rmarkdownfootnote\footnote%
\def\footnote{\protect\rmarkdownfootnote}

\usepackage{titling}

  \title{Assessing the Impact of Gamification on Self-Directed Learning in Medical Students}
    \pretitle{\vspace{\droptitle}\centering\huge}
  \posttitle{\par}
    \author{Lee De Zhang, Vik Gopal, Chan Jia Min, Ng Li Shia and Ang Eng Tat}
    \preauthor{\centering\large\emph}
  \postauthor{\par}
      \predate{\centering\large\emph}
  \postdate{\par}
    \date{}

\begin{document}
\maketitle

{
\setcounter{tocdepth}{2}
}
\section{Introduction}\label{introduction}

``Gamification'' is a relatively new word in the lexicon of the English
language. The Merriam-Webster Dictionary defines it as `the process of
adding games or gamelike elements to something (such as a task) so as to
encourage participation'. In the context of our study, we use the word
to refer to the process of adding game elements to a pedagogical
setting. It is important to note the difference between gamification and
educational or serious games. As pointed out in \cite{dicheva2015gamification},
the latter refers to fully-fledged games for non entertainment purposes,
while gamification merely adds elements of games to an existing process.

Early this millenium, \cite{howarth2002can} mooted the idea
of using games to ``lighten up'' medical education. They observed that
the participation rates for their lunchtime medical quizzes and debates
were higher than those for the regular professional seminars in the
hospital. Theirs was a short report, which was published alongside a
cartoon. Since then, the concept of gamification in education has been
taken up much more seriously. There has been an increasing number of
peer-reviewed papers on this topic. The authors in \cite{nah2014gamification}
selected and discussed 15 peer-reviewed papers on the various
implementations of gamification in education. It was found that these
papers only reported on the game design elements used - little emphasis
was given to evaluating the impact on students' learning. Another review
paper, \cite{hamari2014does}, provides further evidence of
this increased interest in gamification. The authors describe an
increase in the number of peer reviewed papers containing the keywords
``gamification'', ``gamif*" ,``gameful'' and ``motivational affordance''
in various databases. They went on to review 24 of those papers.

The effects of gamification have been studied from a few different
angles. One approach is to measure the success of gamification through
the grades obtained at the end of the semester. However, as suggested in
\cite{muntasir2015gamification}, this might not always be appropriate. Even if we
can establish that the games have led to an increased desire to learn,
we would have to convince the reader that this desire has a causal
effect on grade improvement.

Another approach is to directly correlate the game performance with the
specific skills it is meant to train. In such cases, it is important to
validate the game tool. For instance, by 2012, there had been 25
peer-reviewed papers on the use of actual video games for training
medical and surgical skills. The meta-analysis paper \cite{graafland2012systematic} on this topic focused on the validity of
the games to enhance the learning for participants; it found that 15 of
the 25 tested the validity of the models.

Yet another approach is to measure the impact of the games on the
behavioural and/or psychological outcomes in students. One paper that
performs an excellent analysis in this manner is \cite{beylefeld2007gaming}. They administer a questionnaire to the same group of students
before and after a course that contains game elements. They found that
not only had the games improved students' skills and mastery of the
course content, but they had also led to an improved perception and
attitude towards the course subject.

This paper is a follow up to \cite{tat2018gamifying}. It contains a
quantitative assessment of gamification on the learning values embedded
in medical students. The study in the parent paper was centered on a
validated survey instrument that was utilised to measure the degree of
self-directed learning in students before \textbf{and} after a learning
journey. In this paper, we re-analyse the same study, but with
additional data and with more sophisticated statistical tools. Here, we
apply a PLS path model to the data instead of ANOVA and \(t\)-tests. In
addition, while the original paper addressed the impact of the games on
the grades of the students, we focus here on understanding the impact of
the games on the learning behaviour of the students. To do so, we
capitalise on the clever design of the study, which measures intangible
constructs related to learning before and after the gamification.
Finally, we underline that the analysis in this paper is much more
exploratory - it aids in understanding the students' point of view, and
in hypotheses generation for future studies on gamification.

The remainder of this paper is organised as follows. Section 2 details
the study protocol and contains a brief outline of the statistical
methods used. Section 3 summarises the data and then describes the
output of the statistical models. Finally section 4 contains a
discussion of the results. It raises some interesting observations that
could be further studied in more focused experiments.

\section{Methods}\label{methods}

\subsection{Study Design}\label{study-design}

The study was conducted within the School of Medicine at the National
University of Singapore (NUS). At NUS, classes are taught through a
lecture-tutorial system. Whereas lectures are 1.5 hours long and are
delivered to large groups of students, tutorials are typically 45
minutes long, and consist of small groups of students. Four tutorial
groups of first year medical students were identified for this study.
The four groups of students experienced different levels of game
elements in their Anatomy tutorials. All groups were taught by Dr Ang
Eng Tat; however, they differed in terms of Anatomy topics that they
covered.

\textbf{Group 0} consisted of 16 students from the first semester of the
academic year 2017/2018. Their tutorial classes contained no game
elements.

\textbf{Group 1} consisted of 15 students from the first semester of
academic year 2016/2017. Their tutorials contained games that were meant
to promote healthy competition among individuals and teams. One example
of a team game played was Taboo, where participants had to act out
Anatomy terms to their team-mates.

\textbf{Group 2} consisted of 23 students from the same semester as
Group 1. In addition to the games that Group 1 played, this tutorial
group also had to complete quiz questions on a mobile phone application,
\emph{Teach Me Anatomy}.

\textbf{Group 3} consisted of 22 students from the same semester as
Groups 1 and 2. In addition to the activities that Group 2 participated
in, this tutorial group was also assigned Script Concordance Test (SCT)
questions. SCT questions are designed to enhance clinical reasoning
rather than rote memory. The SCT questions used in this study were
designed by Dr Ng Li Shia, based on the guidelines in \cite{fournier2008script} and \cite{lubarsky2013script}.

In total there were 76 students in the study. The level of gamification
increased monotonically, starting from no games for Group 0 to the most
amount of gamification for Group 3. To provide a tangible motivation to
participate in the activities, there was a SGD 100 prize for the overall
game winner in each of the three groups 1, 2 and 3. Another difference
with Group 0 was that the tutorials for these three groups were
administered as a flipped classroom. This particular implementation
required students to prepare and present an assigned Anatomy topic to
their fellow classmates.

All students were recruited for the study only after obtaining their
informed consent. There were no penalties for withdrawing from the
IRB-approved project (NUS IRB: B-16-205). All students in these four
groups were asked to complete a survey at the beginning and the end of
the semester. The survey questions, further described in Section 2.2,
were \emph{exactly} the same at both points in time.

\subsection{Self-Directed Learning
Questionnaire}\label{self-directed-learning-questionnaire}

A self-directed student is defined as one who takes primary
responsibility or initiative in the learning experience. Medical
practitioners are expected to practice self-directed learning, due to
the rapidly changing nature of requisite clinical knowledge. However,
\cite{tagawa2008physician} noted that the level of self-directed readiness among
medical students is low, and suggested that changes to the medical
school curricula may be able to fix this.

The survey utilised in this study was the Personal Responsibility
Orientation to Self-Direction in Learning (PRO-SDLS) survey. It is
described in full detail in \cite{stockdale2011development}, where the
authors demonstrate the validity of the instrument. There were 25
questions in the survey. Each question was tied to one of four
learning-related constructs: Motivation (7 questions), Initiative (6),
Control (6) and Self-Efficacy (6). The full set of questions can be
viewed in Supplementary Materials I.

The type of Motivation considered here is the type that supports
self-direction: it arises from identifying with the value of the
activity (learning Anatomy) and it arises intrinsically, out of interest
or enjoyment in the activity. Initiative assesses how proactive a
student is with regard to learning. The Control construct is an
indication of how strongly a student feels he can change or influence
his environment in order to learn better. Finally, Self-Efficacy relates
to how confident a student is in his or her own abilities to do what
needs to be done to learn well. This instrument was deemed to be
applicable to adult learners by psychometric experts in  \cite{stockdale2011development}.

Each question in the survey elicited a Likert scale response from the
student. There were 5 possible responses to each question, ranging from
Strongly Disagree to Strongly Agree. Depending on the question, the
response translated to a score from 1 to 5 for that construct. A larger
score indicated a higher value for that learning behaviour.

In our study, the survey was used to assess whether or not the games
increased the level of self-directedness in learning. By administering
it twice - once before the gaming and once after, we are able to pair up
observations for a more powerful model, and then assess if there has
been a change in the attitudes of students.

\subsection{Path Model Analysis}\label{path-model-analysis}

To analyse the data, we employ a Partial Least Squares Path Model (PLS-PM).
This framework allows us to model relationships between blocks of multiple
variables, where each block represents a theoretical construct that is
\textbf{directly unobservable}. In this section, we provide an overview of the
main terms and concepts in path modeling. For further details on path models,
the reader is referred to \cite{sanchez2013pls} and \cite{hair2016primer}.

A path model consists of an inner and an outer model. The inner model
represents the relationships between the latent constructs in our experiment.
In our data, the inner model theorises that the amount of gamification in a
classroom has an impact on the change in Motivation, Initiative, Control and
Self-Efficacy of students. In this paper, we focus on the four latent variables
to understand the impact of the games. A simple visualisation of the inner
model in our analysis can be seen on the left in Figure 2.

Having discussed the inner model, we now turn to the outer model. This
portion defines how the latent variables are uncovered. Latent variables
cannot be directly observed, but they can be indirectly assessed using
instruments such as surveys, or other indicators. These are called
measurement variables, and they can relate to latent variables in one of
two ways: as a \textbf{formative} measurement or as a
\textbf{reflective} one. In the formative case, the measurements define
the latent variable values. In our set-up, the tutorial groups are
formative since they determine the level of games that each student
encounters. On the other hand, the change in score for each question
\textbf{reflects} the value of the latent variable. Hence the changes in
score that we compute from the PRO-SDLS survey are treated as reflective
measurements.

When a PLS-PM is fit to a dataset, the values of the latent variables
are estimated as a linear combination of the measured variables. The
weights in this linear combination are an important part of the output
of a path model. Another important output is the path coefficient on
each arrow. This summarises the strength of the surmised relationship
between two variables. latent variables and the measured variables.
Finally, the goodness of fit of a PLS-PM is assessed using an \(R^2\)
value, similar to that in linear regression.

The estimation procedure for path models requires that the measured
variables be represented numerically. Our experimental set-up contains
four groups that are typically represented as dummy variables. In path
model analysis, there are two main techniques of handling group
variables. One method is to use a permutation test to assess the
importance of the grouping variable. The other is to study the grouping
variable as a mediating or moderating effect. These two approaches are
described further in \cite{chin2004multi} and \cite{dibbern2010introduction}. In
our study, we are exploring the effects of the games, which are employed
at different levels of intensity in the groups. Hence, instead of dummy
variables or as a mediating variable, we represent the groups 0, 1, 2
and 3 with just their integers, reflecting the increase in gamification
from group to group.

\subsection{Principal Component Analysis}\label{pca_intro}

Besides the change in construct levels, another factor of interest is the 
spread (variability) of responses by students. A small spread would indicate
consensus among students regarding the effect of gamification on the constructs
measured, while a large spread would mean little or no consensus. 

However, it would be hard to analyze this by looking at the individual questions 
from each construct, as there are too many questions to consider collectively. 

One way to overcome this is by performing Principal Component Analysis (PCA)  
on the questions belonging to a particular construct. 

PCA reduces the number of variables to be analyzed by building a smaller number of
new variables (known as Principal Components) based on the existing questions. Principal
Components aim to collectively describe the total variance of the data. 

Each Principal Component is ranked based on the amount of variation it explains. The first
Principal Component (PC 1) explains the most variation. In general, the $k^{th}$ Principal 
Component (PC $k$) is the Principal Component which explains the $k^{th}$ most variation. 

The first few Principal Components should ideally explain most of the variation in the dataset. 
If this is the case, differences or similarities within the data can be observed by studying 
these principal components. This is a much easier task to accomplish, as there are fewer 
Principal Components compared to the total number of variables in the original data. 

For a further and more technical explaination of PCA, 
the reader is directed to the book \cite{jolliffe2002} 
and Chapter 10.2 of \cite{james2013introduction}. 

\subsection{Software}\label{software}

Throughout this analysis, we utilise the open-source software R that is
provided by folks at \cite{rsoft2018}, and the package \texttt{plspm}
for path analysis, documented in \cite{plspm2017gaston}.

\section{Results}\label{results}

\subsection{Exploratory Findings}\label{exploratory-findings}

\begin{figure}[H]
\includegraphics[width=0.5\linewidth]{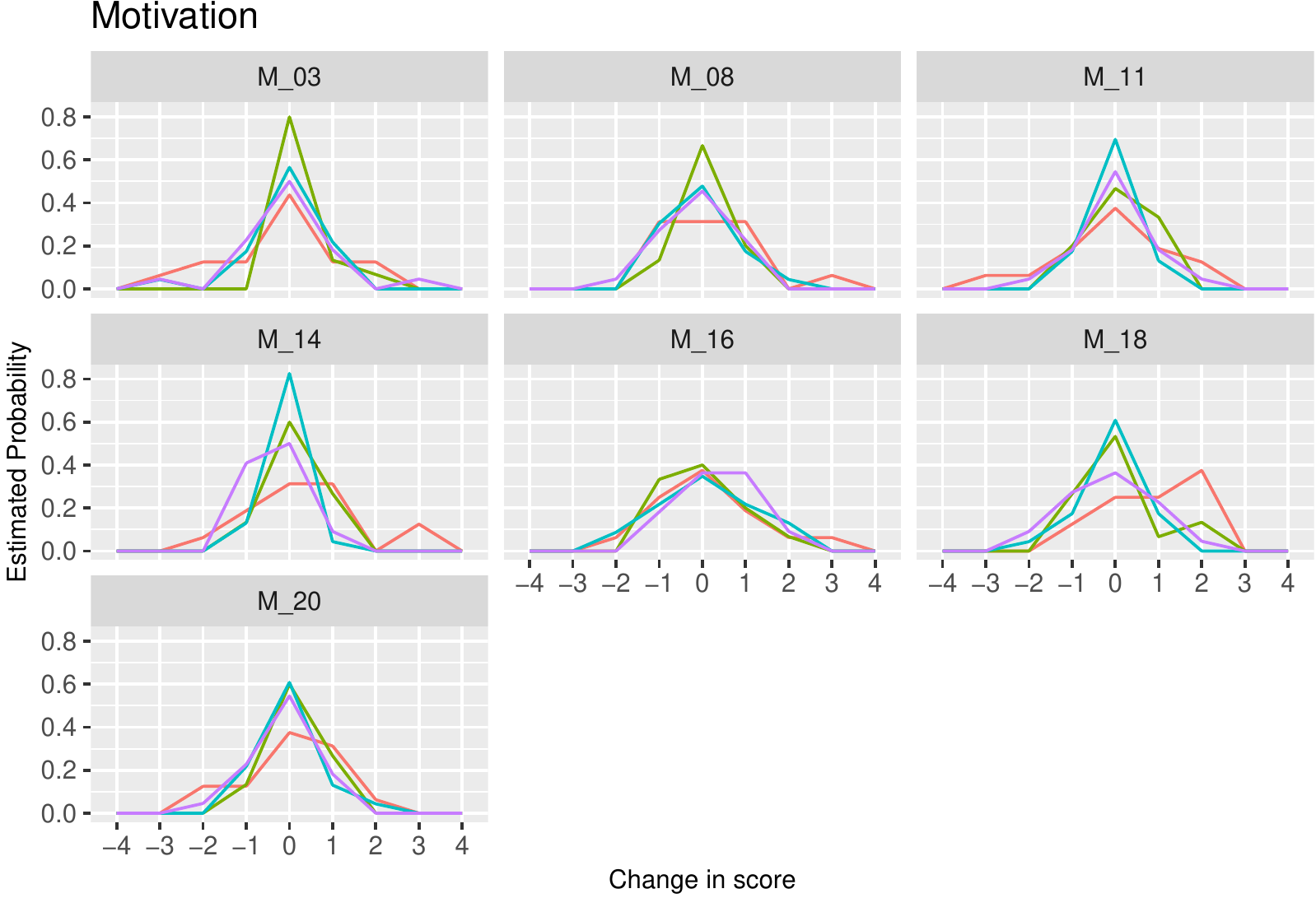}
\includegraphics[width=0.5\linewidth]{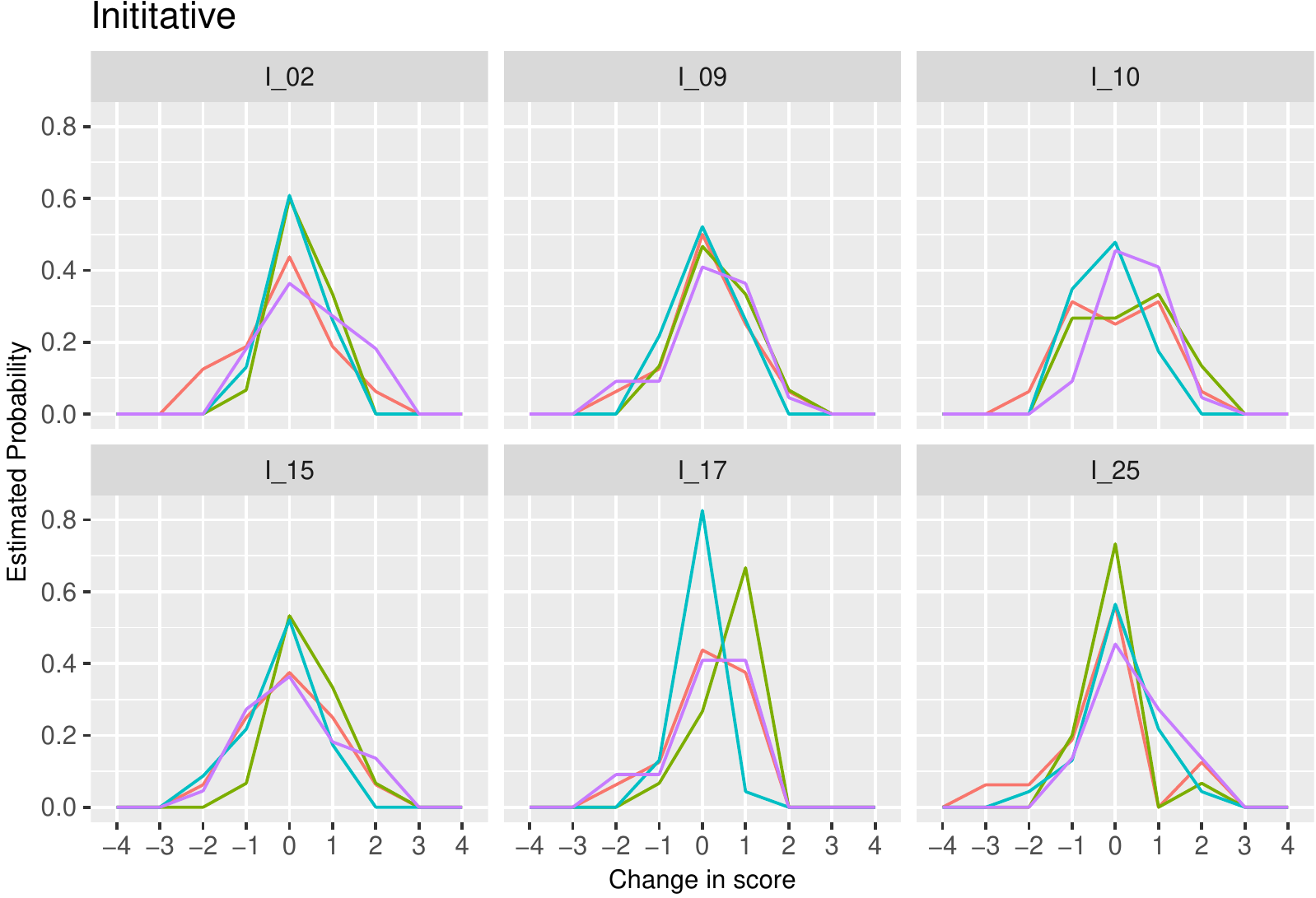}
\includegraphics[width=0.5\linewidth]{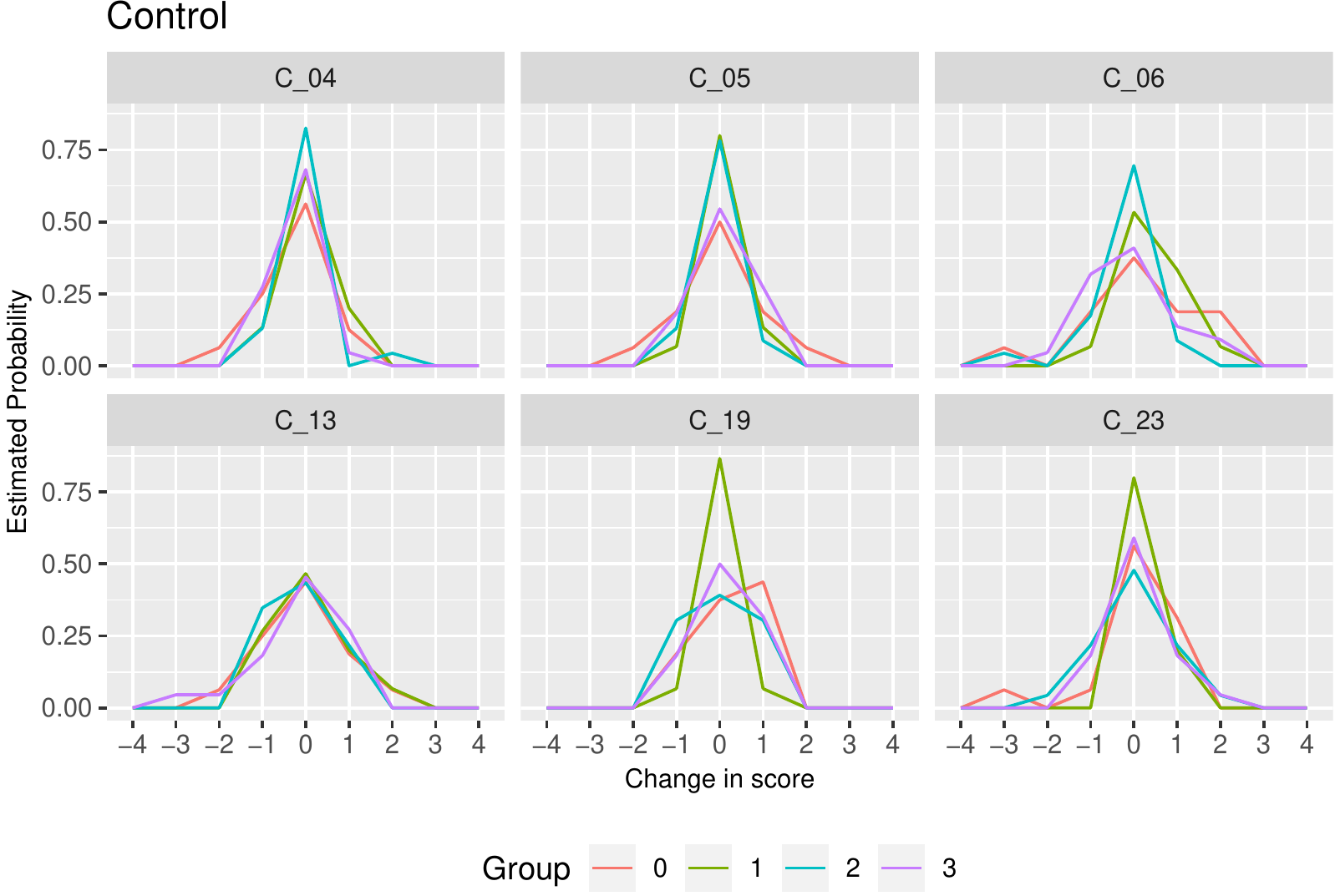}
\includegraphics[width=0.5\linewidth]{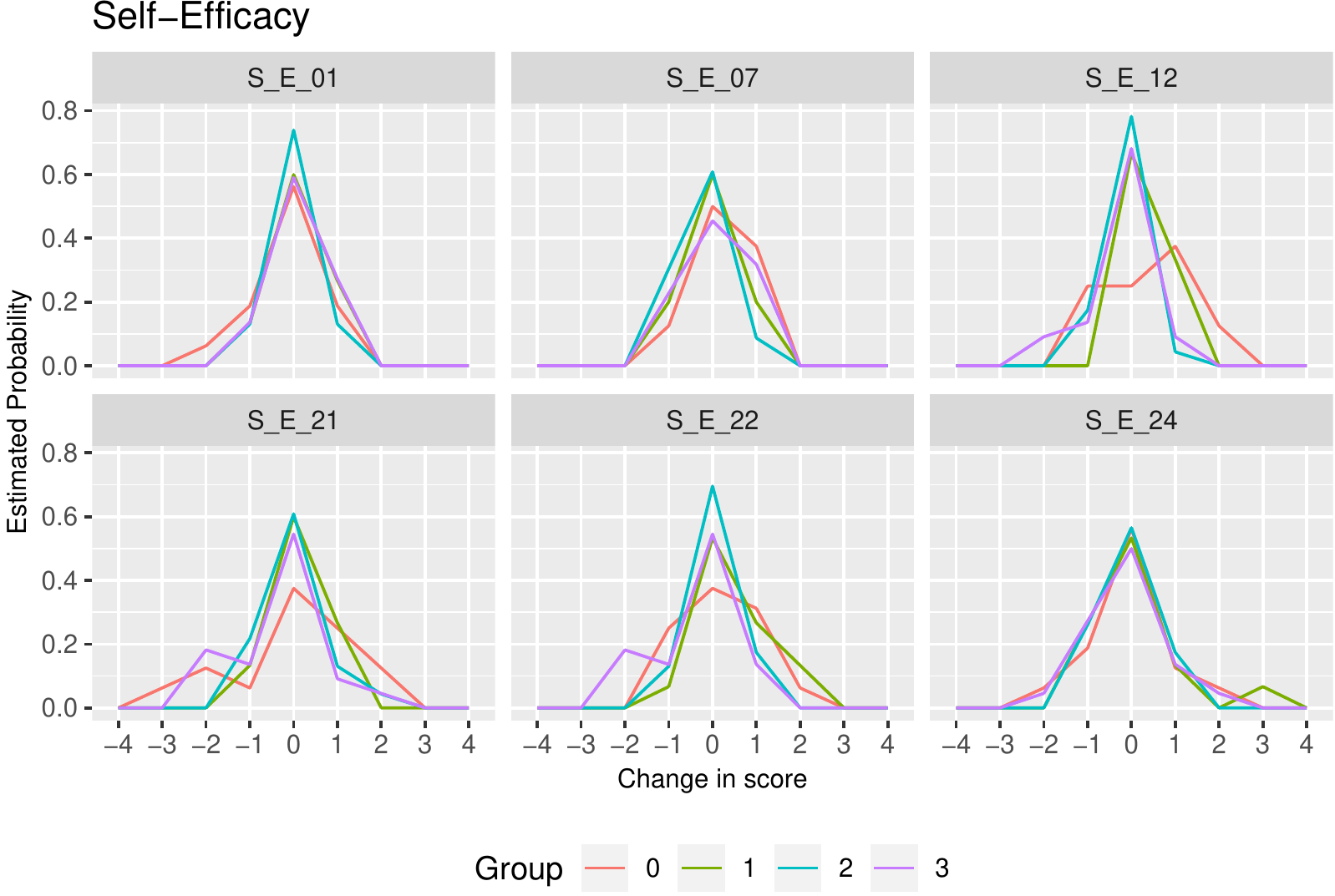}
\caption{Estimated distribution of change in responses for each question
answered by each group.}
\label{fig:pmfs}
\end{figure}

In order to get a grasp on the main patterns in the data, we begin by
subtracting the pre-feedback score for each question from the post-feedback
score. With these 25 values for each individual, we estimate a probability mass
function for each group within each question. The results are presented in
Figure \ref{fig:pmfs}. Most of the panels display a symmetric tent shape,
centred at 0, with little difference between the groups. This indicates that
the games had little effect on the students' responses for these questions.

However, we invite the viewer to take a close look at questions 14, 18
(under Motivation), 6 (under Control), 12, and 22 (under Self-Efficacy).
For each of these panels, we can see that the skewness of the distributions
changes from left to right as we consider the red, green, blue and then
purple lines. This corresponds to increases in construct scores for
these questions as we traverse the groups from 0 to 3, i.e.~in order of
\textbf{decreasing} gamification. On the other hand, if we study the
panels for Initiative, we might deduce the trend was in the opposite
direction. 

Since we are interested in whether a student's score \emph{decreased, remained
the same} or \emph{increased}, we work out these proportions for each
construct and display them in Table \ref{tab:props}.

\begin{table}[ht]
\centering
\begin{tabular}{r|rrr|rrr}
  \hline
  & \multicolumn{3}{|c}{Motivation} & \multicolumn{3}{|c}{Initiative} \\
  \hline
Group & Decreased & Same & Increased & Decreased & Same & Increased \\ 
  \hline
  0 & 0.27 & 0.35 & \textcolor{red}{0.38} & 0.28 & 0.43 & 0.29 \\ 
    1 & 0.17 & 0.58 & 0.25 & 0.13 & 0.48 & \textcolor{red}{0.39} \\ 
    2 & 0.22 & 0.59 & 0.19 & 0.22 & 0.59 & 0.20 \\ 
    3 & 0.29 & 0.47 & 0.24 & 0.18 & 0.41 & \textcolor{red}{0.41} \\ 
   \hline
\end{tabular}
\end{table}\begin{table}[ht]
\centering
\begin{tabular}{r|rrr|rrr}
  \hline
  & \multicolumn{3}{|c}{Control} & \multicolumn{3}{|c}{Self-Efficacy} \\
  \hline
Group & Decreased & Same & Increased & Decreased & Same & Increased \\ 
  \hline
  0 & 0.24 & 0.47 & 0.29 & 0.23 & 0.44 & 0.33 \\ 
    1 & 0.10 & 0.69 & 0.21 & 0.13 & 0.59 & 0.28 \\ 
    2 & 0.23 & 0.60 & 0.17 & 0.20 & 0.67 & 0.13 \\ 
    3 & 0.24 & 0.53 & 0.23 & 0.26 & 0.55 & 0.19 \\ 
   \hline
\end{tabular}
\caption{Crude proportions of change for each construct.}
\label{tab:props}
\end{table}

Once again, we observe that the category of no change always has the
highest proportion of cases. In addition, the proportion of positive
changes appears to decrease as we go down the table (increasing
gamification) for Motivation, Control and Self-Efficacy. Within these
three, this trend appears strongest for Motivation and weakest for
Control. Although it must be noted that there is no corresponding
dramatic uptick for the proportion of negative changes for these groups.
The Initiative construct once again bucks this trend, with group 3
having the highest proportion of positive changes.

\subsection{Path Model Results}\label{path-model-results}

The PRO-SDLS survey had already been ratified by the original authors
Stockdale and Brockett (2011). They had found that the individual
questions for each construct were consistent. In the output for our
model too, we find that the individual blocks of questions are
unidimensional. In line with general guidelines for path models as
outlined in \cite{hair2016primer}, the Dillon-Goldstein's rho values are
all above 0.7. The goodness-of-fit of the model was 0.17. This indicates
poor predictive value of the model, but, as we shall see in the next
section, there is value in interpreting the coefficients and output in
the context of the study. We obtain comprehensible and reasonable
findings in addition to a host of unanswered questions!

\begin{figure}[H]
\includegraphics[width=0.5\linewidth]{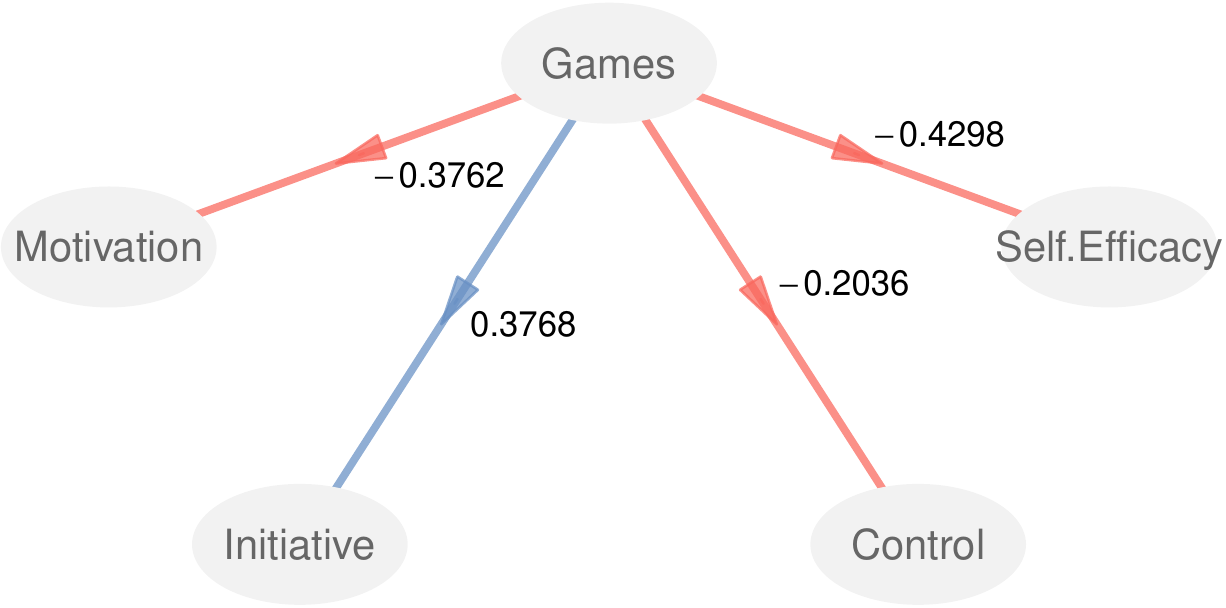}
\includegraphics[width=0.5\linewidth]{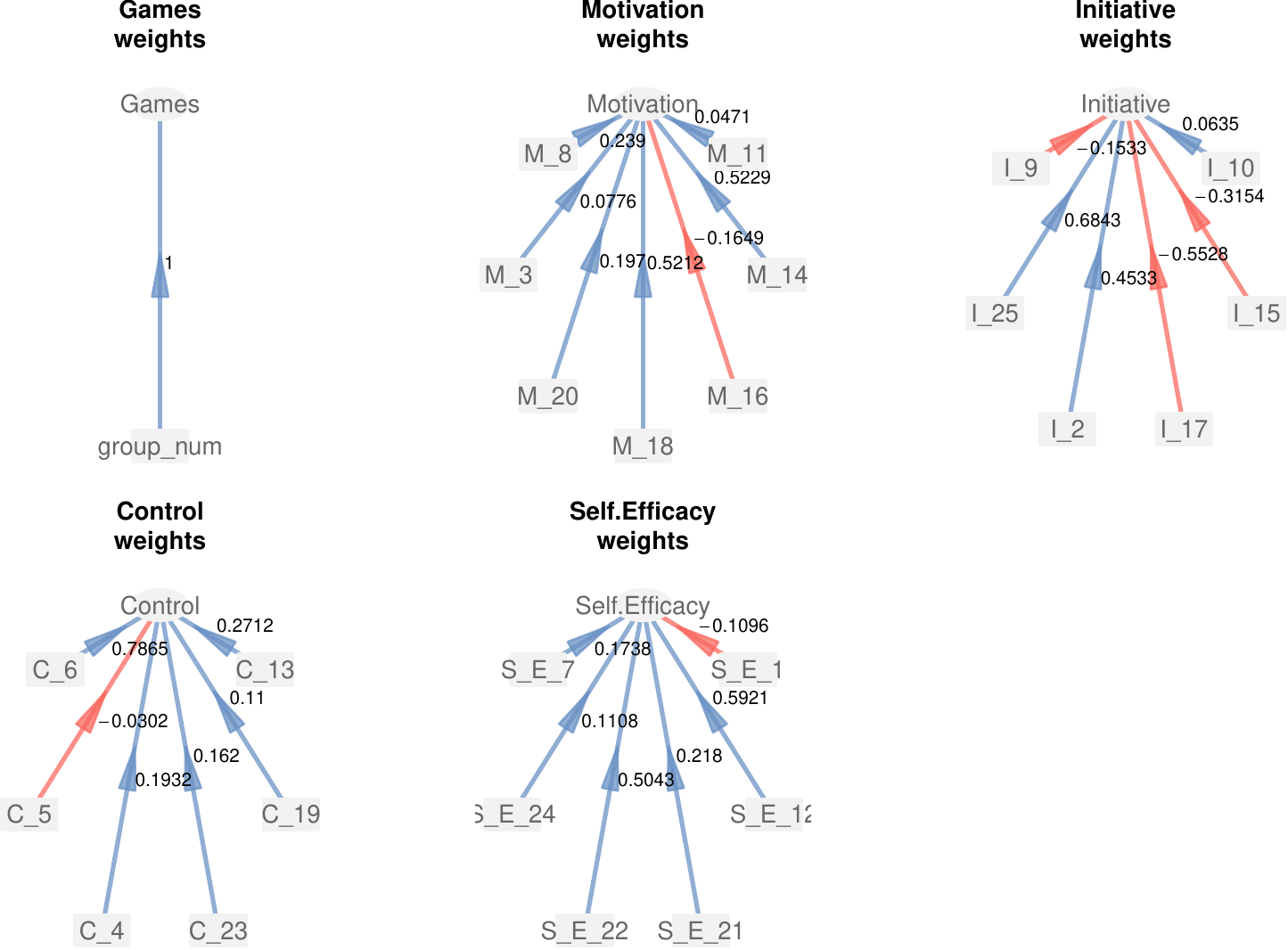}
\caption{Path model output.}
\label{fig:path}
\end{figure}

When we inspect the output of the path model, we find that only the path
coefficient for Self-Efficacy is significantly different from 0: the
95\% bootstrap confidence interval for this parameter is (-0.623,
-0.218). This indicates that as the level of games included increased,
the Self-Efficacy of students significantly decreased. The direction of
the relationship is the same for the Control and Motivation constructs.
It is only for Initiative that the associated change in level is
positive. The inclusion of more games seemed to steer the students
towards becoming proactive. However, the coefficients for the latter
three paths are not significantly different from 0.

Let us recall that the weights are the coefficients that, when applied
to the score changes for individual questions, result in the latent
variable values for each construct. The weights for the measured
variables in the Motivation block are positive except for one question
(14). The scenario is identical to the other two blocks whose path
coefficients are also negative: Control and Self-Efficacy. For the
Initiative block, the weights for half of the measured variables are
positive while the rest are negative. In the next section, we zoom in on
the questions with large weights and the questions within each block
whose signs are in the minority.

\subsection{Principal Component Analysis Results}\label{pca_result}

Principal Component Analysis (PCA) was then carried out on the questions, 
grouped by the construct they were intended to measure. 

\begin{table}[H]

\begin{minipage}{.5\linewidth}
	\begin{subtable}[t]{.3\textwidth}
	\centering
		\begin{tabular}{|r|rrr|}
 		 \hline
		 & PC1 & PC2 & PC3  \\ 
 		 \hline
		Standard deviation & 1.5177 & 1.0664 & 1.0114  \\ 
 		 Proportion of Variance & 0.3291 & 0.1624 & 0.1461  \\ 
 		 Cumulative Proportion & 0.3291 & 0.4915 & 0.6376  \\ 
  			 \hline
		\end{tabular}
	
  	 \caption{Motivation}
	\end{subtable}

	\hfill

	\begin{subtable}[t]{0.3\textwidth}
		\centering
		\begin{tabular}{|r|rrr|}
 		 \hline
		 & PC1 & PC2 & PC3  \\ 
 		 \hline
		Standard deviation & 1.3831 & 1.1462 & 0.9853  \\ 
 		 Proportion of Variance & 0.3188 & 0.2190 & 0.1618 \\ 
 		 Cumulative Proportion & 0.3188 & 0.5378 & 0.6996  \\ 
  		 \hline
  		\end{tabular}
  		\caption{Control}
\end{subtable}
	
\end{minipage}%
\begin{minipage}{.5\linewidth}
	
	\begin{subtable}[t]{.3\textwidth}
		\centering
		\begin{tabular}{|r|rrr|}
		  \hline
		 & PC1 & PC2 & PC3 \\ 
 		 \hline
		Standard deviation & 1.3399 & 1.0727 & 0.9757  \\ 
 		 Proportion of Variance & 0.2992 & 0.1918 & 0.1587  \\ 
 		 Cumulative Proportion & 0.2992 & 0.4910 & 0.6497  \\ 
 		  \hline
		\end{tabular}
	\caption{Initiative}
	\end{subtable}
	
	\hfill
	
	\begin{subtable}[t]{0.3\textwidth}
		\centering
		\begin{tabular}{|r|rrr|}
		  \hline
 			& PC1 & PC2 & PC3 \\ 
		  \hline
		Standard deviation & 1.3624 & 1.1288 & 0.9790 \\ 
 		 Proportion of Variance & 0.3093 & 0.2124 & 0.1597  \\ 
		  Cumulative Proportion & 0.3093 & 0.5217 & 0.6815 \\ 
 		  \hline
		\end{tabular}
	\caption{Self-Efficacy}
	\end{subtable}
	
\end{minipage}

	\caption{ Summary of the first three Principal Components from each construct.
}
\label{tab:pca_summary}
\end{table}
	
Based on Table \ref{tab:pca_summary}, we note that for all the constructs, 
Principal Component 1 (PC 1) and Principal Component 2 (PC 2) cumulatively 
explain at least $49.1\%$ of the total variance. In other words, this means 
that the first two Principal Components for each construct explains around 
half the total variation of the original data. Analyzing both concurrently
should give us a good idea of the spread of our data.

\begin{figure}[H]
        \centering
        \begin{subfigure}[t]{.4\textwidth}
            \centering
            \includegraphics[width=\linewidth]{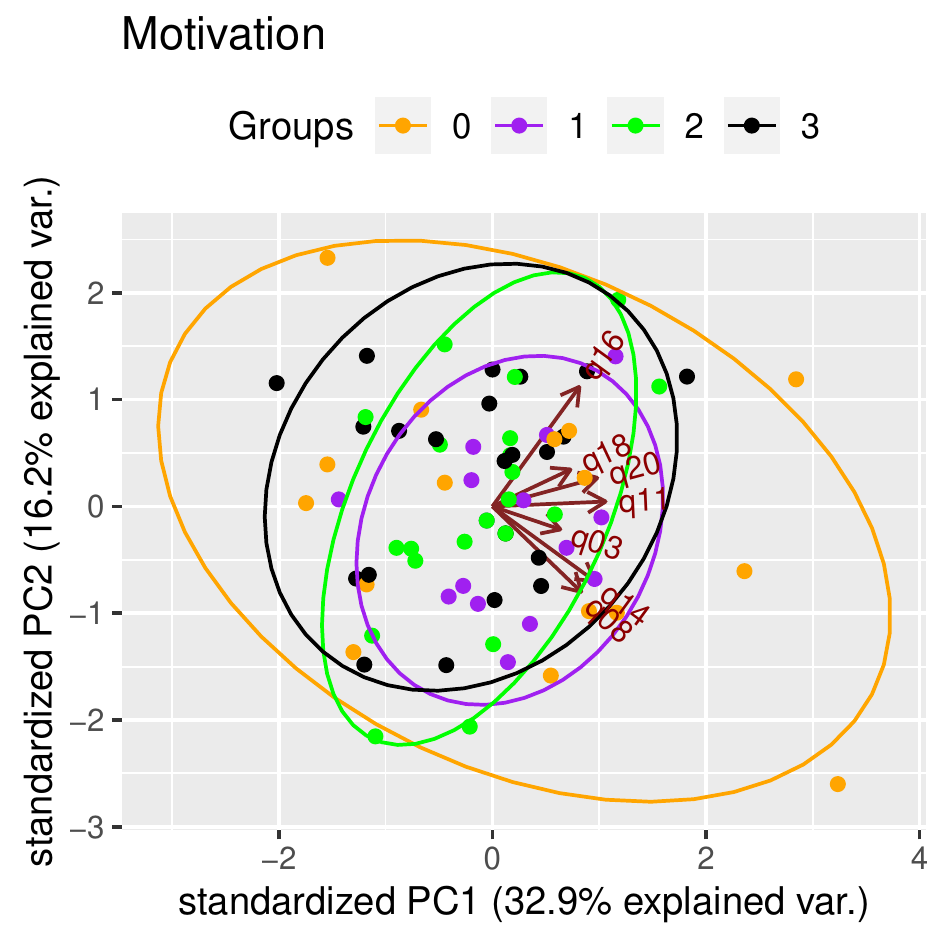}
            \label{fig:se_pca}
        \end{subfigure}
        \hfill
        \begin{subfigure}[t]{.4\textwidth}  
            \centering 
            \includegraphics[width=\linewidth]{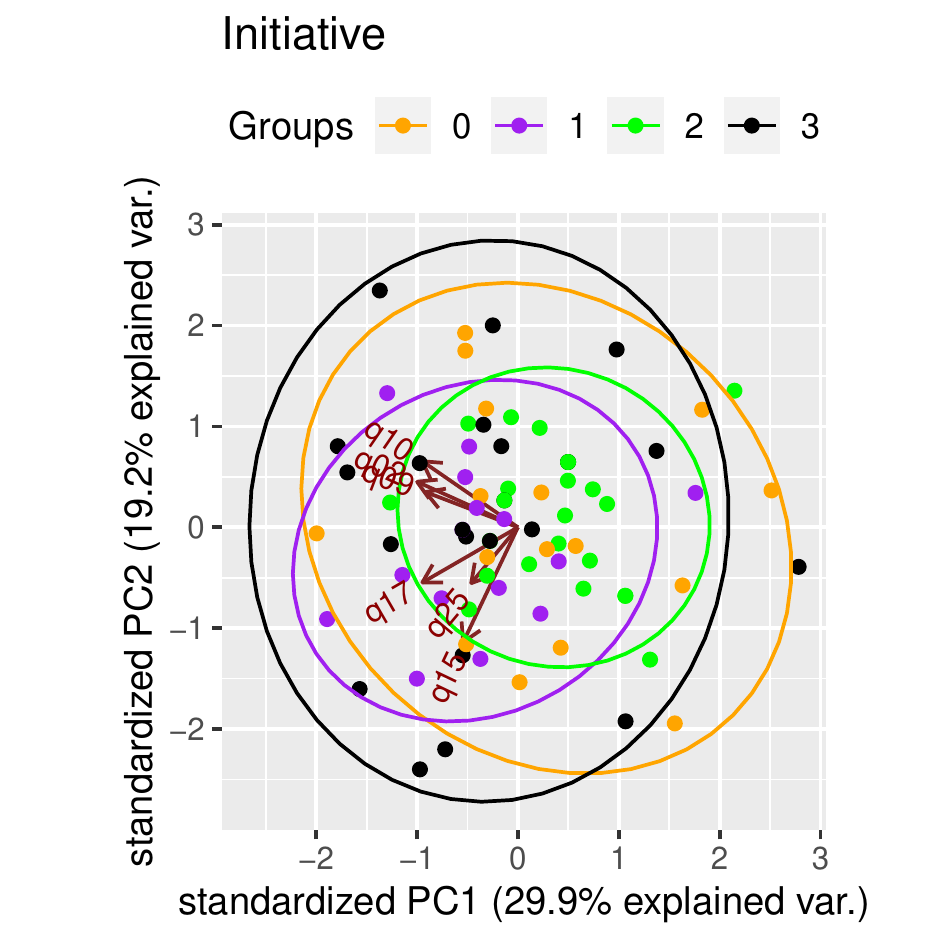}
            \label{fig:ctrl_pca}
        \end{subfigure}
        \medskip
        \begin{subfigure}[t]{.4\textwidth}   
            \centering 
            \includegraphics[width=\linewidth]{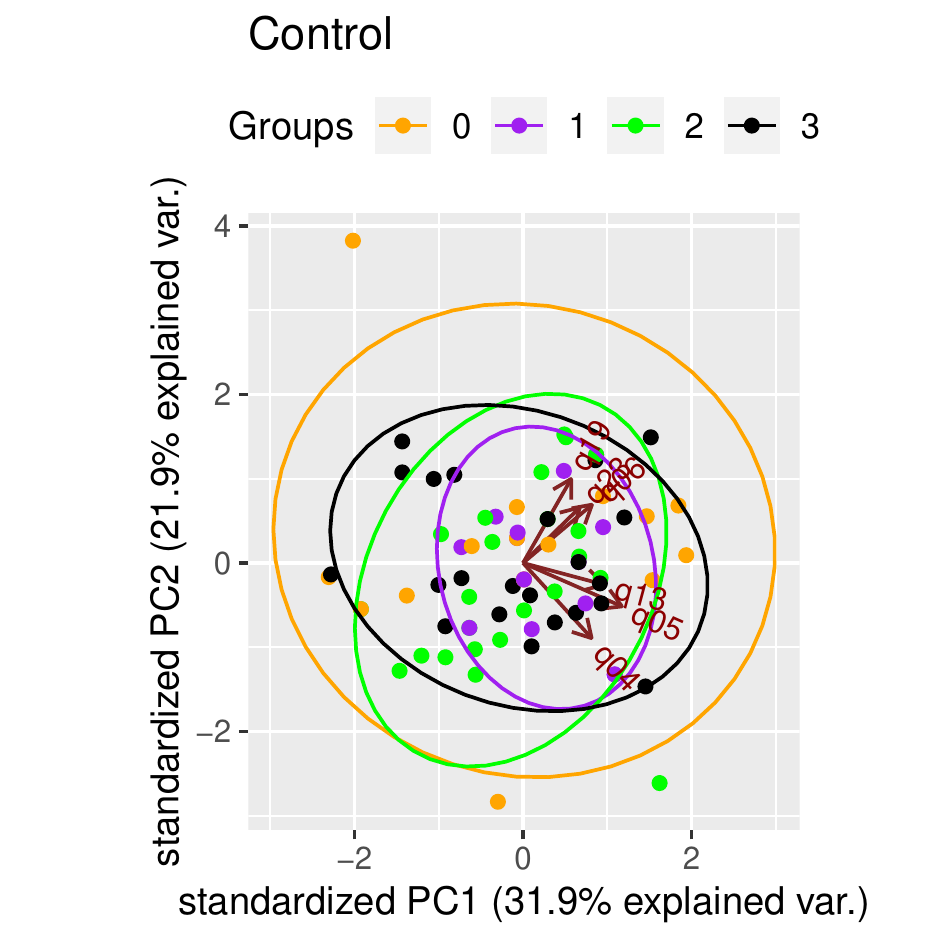}
            \label{fig:init_pca}
        \end{subfigure}
        \hfill
        \begin{subfigure}[t]{.4\textwidth}   
            \centering 
            \includegraphics[width=\linewidth]{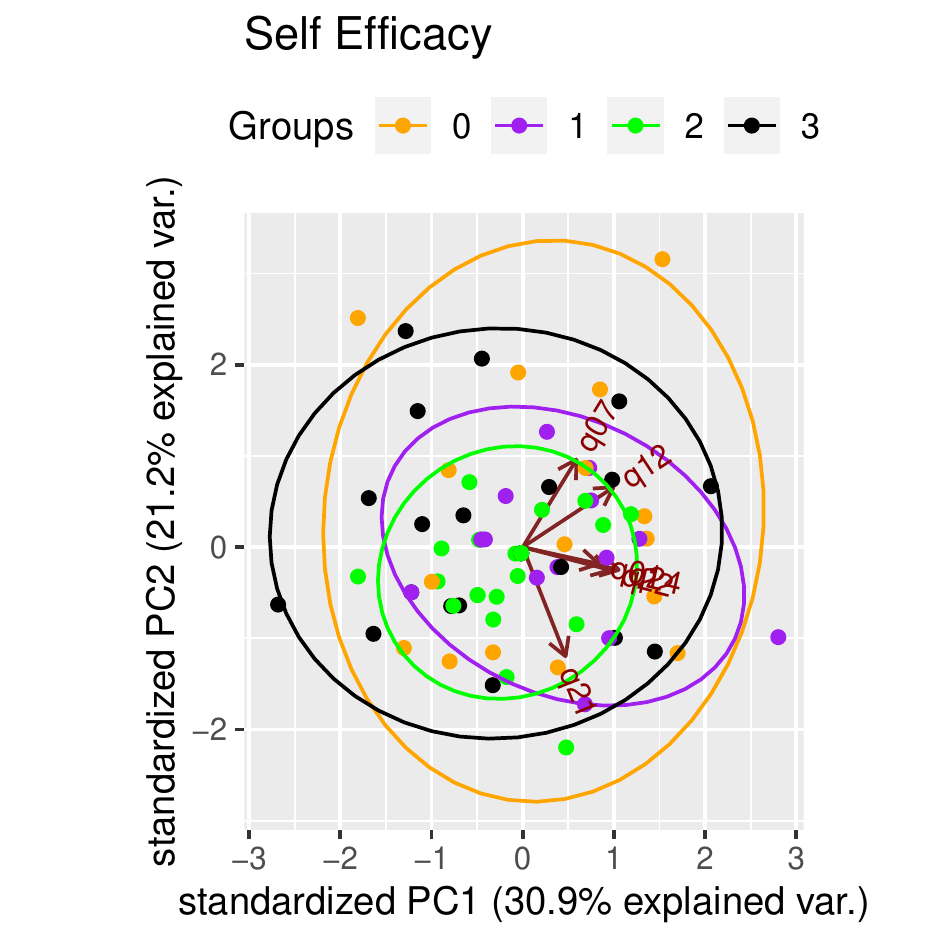} 
            \label{fig:moti_pca}
        \end{subfigure}
        \caption{Plots of Principal Component 2 against Principal Component 1, with the corresponding confidence ellipses, for all constructs.}
        \label{fig:pca_all}
    \end{figure}

For each construct, Principal Component 2 (PC 2),was plotted
 against Principal Component 1 (PC 1). Following which, a $90\%$ confidence ellipse 
was then plotted for each of the groups. A $x\%$ confidence ellipse refers to the 
region where a new observation (which was not used to compute the Principal 
Components) will fall in with a probability of $x\%$. Figure \ref{fig:pca_all}
shows these plots.

The size of confidence ellipse is a result of the variability of the data. 
If the data has a low variance, it means that the data is closely clustered 
and hence, previously unseen observations have a high probability of being 
close to this cluster. As a result, the confidence ellipse will be smaller.

On the other hand, if the data is spread out (high variance), it would be harder 
to predict the position of previously unseen observations. Therefore, the 
confidence ellipse will be bigger to compensate for this uncertainty.

We shall detail our findings based on the PCA plots in the next section.

\section{Discussion}\label{discussion}

\subsection{Interpreting the Path Model
Output}\label{interpreting-the-path-model-output}

In this section, we discuss the interpretation of the path model
coefficients and weights. We summarise the findings in the conclusion
section in Section 4.3.

\subsubsection{Motivation Block}\label{motivation-block}

The path coefficient suggests that Motivation levels decreased as
gamification increased. In line with the exploratory plots, the
questions with the largest weights are 14 and 18:

\begin{itemize}
\tightlist
\item
  14: \emph{Most of the work that I do in my courses is personally
  enjoyable or seems relevant to my reasons for attending university.}
  Increased gamification was associated with stronger levels of
  disagreement with this statement.
\item
  18: \emph{The main reason that I do the course activities is to avoid
  feeling guilty or getting a bad grade.} Increased gamification
  \(\Rightarrow\) stronger agreement.
\end{itemize}

We can only hypothesise, but it suggests that when there were too many
game elements, the students felt it was a distraction. Taking the
phrasing of these two statements into consideration, we are loathe to
conclude that Motivation decreased. In fact, we are almost certain that
our students continued to work hard, but perhaps it was the case that
they found it difficult to believe the game-centric sessions could help
them as much as routine or traditional learning.

\subsubsection{Initiative Block}\label{initiative-block}

Unlike the others', the path coefficient for this block was estimated to
be positive. It meant that on average, increased gamification led to an
increase in this construct. In particular, a relatively large negative
weight was estimated for the following question:

\begin{itemize}
\tightlist
\item
  17: \emph{I often collect additional information about interesting
  topics even after the course has ended.} Increased gamification
  \(\Rightarrow\) stronger agreement.
\end{itemize}

Upon matching this output with the estimated mass functions earlier,
however, we observe that this ``trend'' could be due to group 1 alone -
their answers for this questions were the most changed. They remaining
groups' had not changed much for this question.

The largest positive weight was attributed to question 25:

\begin{itemize}
\tightlist
\item
  25: \emph{I always rely on the professor/lecturer to tell me what I
  need to do in the course to succeed.} Increased gamification
  \(\Rightarrow\) stronger disagreement.
\end{itemize}

On the whole, the painted picture suggests that students drove
themselves into action when they assessed the intensity of non-serious
elements in the classroom.

\subsubsection{Control Block}\label{control-block}

The negative path coefficient for this block was not significant. There
was only one question that had a large weight here:

\begin{itemize}
\tightlist
\item
  6: \emph{I often have a problem motivating myself to learn.} Increased
  gamification \(\Rightarrow\) stronger agreement.
\end{itemize}

Again, it could be that the amount of games led to reduced interest in
the topic.

\subsubsection{Self-Efficacy Block}\label{self-efficacy-block}

This was the only block whose path coefficient was significant. Being
negative, it implied that as gamification increased, the self-confidence
of students fell.

\begin{itemize}
\tightlist
\item
  12: \emph{I am very convinced I have the ability to take personal
  control of my learning.} Increased gamification \(\Rightarrow\)
  stronger disagreement.
\item
  22: \emph{I am unsure about my ability to independently find needed
  outside materials for my courses.} Increased gamification
  \(\Rightarrow\) stronger agreement.
\end{itemize}

The indication seems to be that students are worried by the game
elements; perhaps they feel there is less rigour in the class. Hence
they fear that they will be ill-prepared for their job later on.

\subsection{Interpreting the Principal Component Analysis Plots}\label{pca_discussion}

First, we note that for the self-efficacy, control and motivation constructs, 
the confidence ellipse for Group 0 (without gamification) is the largest. 
For the initiative construct, the confidence ellipse for Groups 0 and 3 are 
roughly the same size, and are significantly larger compared to Groups 1 and 2. 
This indicates that generally, there is a large spread (variance) among 
responses in Group 0 (and for initiative, Group 3) compared to the other groups.

For the self-efficacy and Initiative construct, Group 2 has the smallest
 confidence ellipse compared to the other groups. On the other hand, Group 1 
 has the smallest confidence ellipse for the motivation and control constructs. 
 This means that there is less variability among the difference in 
 pre and post responses for students from these groups.

Overall, we notice that typically, the groups with gamification (Groups 1, 2, 3) 
have a smaller confidence ellipse compared to the group without gamification (Group 0). 
In other words, there is less variation among responses from students belonging 
to the former groups. This is particularly the case for the motivation construct,  
where the ellipse for Group 0 is more than twice the area for the other groups. 

This suggests that there is some form of consensus among groups with gamification, 
resulting in more similar changes in construct intensity pre and post 
intervention compared to the group without gamification. 

Among the groups with gamification, Group 3 has the biggest confidence 
ellipse for all the constructs, compared to Groups 1 and 2. This suggests 
that students' opinions on the effectiveness of the games employed in 
Group 3 is more varied compared to Groups 1 and 2. 

An alternative explanation for the difference in spread across the groups is that the 
students formed their own opinions regarding the effectiveness of the games employed 
collectively as a group. Since this wasn't a controlled experiment, the students were
free to discuss the games with their classmates, and even compare their experiences with
students from other cohorts (where different or no games were employed). 

For example, if a student felt that the games employed was not effective or to his/her 
liking, the student may have voiced this out to his/her classmates. As a result, the opinion
of the classmate may have changed as a result of this and by word of mouth, this may further spread
to other classmates. This would then result in a consensus among all the students about the games.

\subsection{Feedback from Gamified
Groups}\label{feedback-from-gamified-groups}

At the end of the semester, each student in groups 1, 2 and 3 was polled
to obtain qualitiative feedback on the game-enhanced tutorials that they
had participated in. The number of respondents from each of these three
groups was 14, 22 and 19 respectively. The full feedback form is
available for perusal as Supplemental Material II.

All three groups experienced a flipped classroom, meaning that students
were assigned to present topics to their peers during the tutorial
sessions. This was a success with the students. The proportion of
students who answered yes to the following question was 1, .77 and .97
for the three groups.

\begin{quote}
Did the `flipped classroom' motivate you to do your own self-directed
learning?
\end{quote}

One of the free-text questions in the form queried the students as to
which part of the tutorial sessions they found most useful for learning.
The wordcloud for it, below, indicates that most of them found the
flipped classroom aspect most useful. The words \emph{presentation},
\emph{peer/classmates learning} and \emph{flipped} appear repeatedly.

\begin{figure}[H]
\centering
\includegraphics{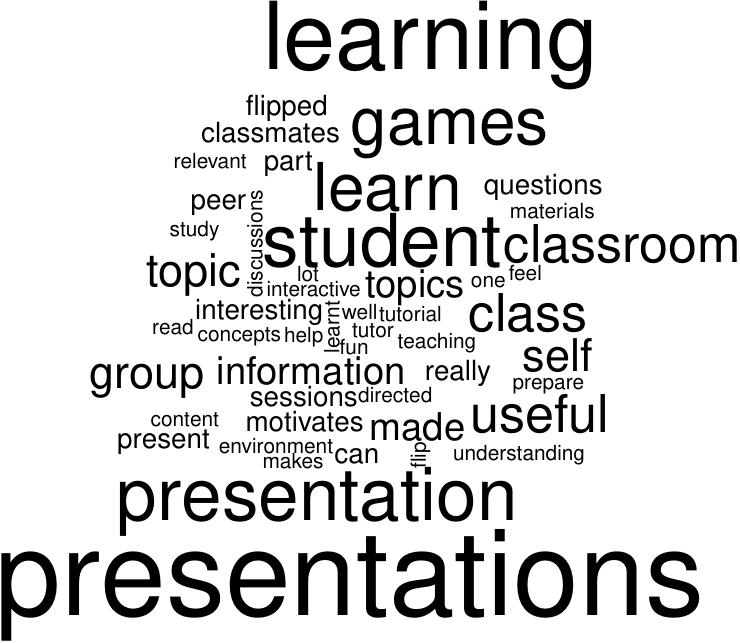}
\caption{Wordcloud for free-text question on useful elements of the class.}
\label{fig:wordcloud}
\end{figure}

It is interesting to note that, corresponding to an increase in amount
of games, there was a decreasing proportion of students who felt that
the in-class games should continue. The proportion of such students fell
from 1.00 to 0.82 to 0.76, corresponding to groups 1, 2 and 3. The
precise wording of the question was:

\begin{quote}
Should this type of tutorial-games sessions continue in future in place
of conventional tutorials?
\end{quote}

Perhaps the groups with more gamification found the more serious game
elements, namely the mobile review app and the Script Concordance Test,
more beneficial than the in-class games that were played. However, based
on the numbers alone, the serious game elements were not unanimously
acknowledged as useful either. The proportion of group 3 students who
found the SCT useful was 0.71, although it was explicitly mentioned by
two of the students as the most motivating part of the sessions. The
proportion of group 2 and 3 students who found the mobile app useful was
only 0.36 (in both cases). Overall, the Final Prize awarded to an
individual did not appear to be a motivating factor for the students. In
each group, the proportion of students who thought so was .50 or less
(.50, .23, and .47 respectively).

Although the path model indicates a drop in Motivation to go with an
increase in gamification, one of the free-text questions seems to
contradict it.

\begin{quote}
Overall, did you feel more motivated to do self-directed learning after
this series of tutorial-games sessions? How so?
\end{quote}

Overall, only 12 out of the 55 respondents included a \emph{no} or
\emph{not really} in their answer. Slightly more than half (30 out of
55) responded with a \emph{yes} to the question above.

\subsection{Conclusions}\label{conclusions}

To summarise, we have found that introducing game elements into
tutorials needs to be done with care. Although the models that we have
fitted do not have highly significant results, they have provided great
insights into the impact of gamification on students. We should bear in
mind that we do not claim the above results are reproducible. For
instance, they could be esoteric to Asian students, who are typically
already very well motivated to study on their own, and to study hard.

One finding was that Self-Efficacy fell significantly as the level of
games increased. If students truly are concerned about material being
sacrificed for the purpose of gamification, then we need to consider how
gamification studies should be introduced in future studies. A study
where two groups (one gamified and one not) are compared via their
performance on an oral exam could be introduced. Alternatively, video
lectures from previous semesters could be released to students to
alleviate their concerns about the quantity of material they cover.

Regarding Motivation, the path model suggests that the students did not
find the games appealing. Although a slight majority in the free-text
question answered that the games motivated them, the feedback on the
Final Prize agrees with the path model. This is an undesirable
situation, as outlined in \cite{muntasir2015gamification}. There, the authors
point out that the rewards in the gamification should resonate with the
value of the user. One possibility might be to include the games as a
small component in the overall grade of the course.

The qualitative feedback also revealed the value of the flipped
classroom. Students were able to appreciate the value of the
presentations and the process of preparing for them.

The final two points are to highlight the design of the study, and the
use of a path model to analyse the survey data. The before-after set up
allowed for powerful comparisons to be made, and the path model provided
an easily interpretable model for discussion.

\bibliography{ref}

\begin{thebibliography}{10}

\bibitem{dicheva2015gamification}
Dicheva D, Dichev C, Agre G, Angelova G.
\newblock Gamification in education: A systematic mapping study.
\newblock Journal of Educational Technology \& Society. 2015;18(3).

\bibitem{howarth2002can}
Howarth-Hockey G, Stride P.
\newblock Can medical education be fun as well as educational?
\newblock BMJ: British Medical Journal. 2002;325(7378):1453.

\bibitem{nah2014gamification}
Nah FFH, Zeng Q, Telaprolu VR, Ayyappa AP, Eschenbrenner B.
\newblock Gamification of education: a review of literature.
\newblock In: International conference on hci in business. Springer; 2014. p.
  401--409.

\bibitem{hamari2014does}
Hamari J, Koivisto J, Sarsa H.
\newblock Does gamification work?--a literature review of empirical studies on
  gamification.
\newblock In: 2014 47th Hawaii international conference on system sciences
  (HICSS). IEEE; 2014. p. 3025--3034.

\bibitem{muntasir2015gamification}
Muntasir M, Franka M, Atalla B, Siddiqui S, Mughal U, Hossain IT.
\newblock The gamification of medical education: a broader perspective.
\newblock Medical education online. 2015;20.

\bibitem{graafland2012systematic}
Graafland M, Schraagen JM, Schijven MP.
\newblock Systematic review of serious games for medical education and surgical
  skills training.
\newblock British journal of surgery. 2012;99(10):1322--1330.

\bibitem{beylefeld2007gaming}
Beylefeld AA, Struwig MC.
\newblock A gaming approach to learning medical microbiology: students’
  experiences of flow.
\newblock Medical teacher. 2007;29(9-10):933--940.

\bibitem{tat2018gamifying}
Ang ET, Min CJ, Gopal V, Li~Shia N.
\newblock Gamifying Anatomy Education.
\newblock Clinical Anatomy. 2018;.

\bibitem{fournier2008script}
Fournier JP, Demeester A, Charlin B.
\newblock Script concordance tests: guidelines for construction.
\newblock BMC medical informatics and decision making. 2008;8(1):18.

\bibitem{lubarsky2013script}
Lubarsky S, Dory V, Duggan P, Gagnon R, Charlin B.
\newblock Script concordance testing: From theory to practice: AMEE Guide No.
  75.
\newblock Medical teacher. 2013;35(3):184--193.

\bibitem{tagawa2008physician}
Tagawa M.
\newblock Physician self-directed learning and education.
\newblock The Kaohsiung journal of medical sciences. 2008;24(7):380--385.

\bibitem{stockdale2011development}
Stockdale SL, Brockett RG.
\newblock Development of the PRO-SDLS: A measure of self-direction in learning
  based on the personal responsibility orientation model.
\newblock Adult Education Quarterly. 2011;61(2):161--180.

\bibitem{sanchez2013pls}
Sanchez G.
\newblock PLS path modeling with R.
\newblock Berkeley: Trowchez Editions. 2013;.

\bibitem{hair2016primer}
Hair~Jr JF, Hult GTM, Ringle C, Sarstedt M.
\newblock A primer on partial least squares structural equation modeling
  (PLS-SEM).
\newblock Sage Publications; 2016.

\bibitem{chin2004multi}
Chin WW.
\newblock Multi-group analysis with PLS.
\newblock Frequently asked questions-partial least squares \& PLS-graph. 2004;.

\bibitem{dibbern2010introduction}
Dibbern J, Chin W.
\newblock An introduction to a permutation based procedure for multi-group PLS
  analysis: results of tests of differences on simulated data and a cross
  cultural analysis of the sourcing of information system services between
  Germany and the USA.
\newblock Handbook of Partial Least Squares: Concepts, Methods and
  Applications. 2010;p. 171--193.

\bibitem{jolliffe2002}
Jolliffe IT.
\newblock Principal Component Analysis.
\newblock Springer; 2002.

\bibitem{james2013introduction}
James G, Witten D, Hastie T, Tibshirani R.
\newblock An introduction to statistical learning. vol. 112.
\newblock Springer; 2013.

\bibitem{rsoft2018}
{R Core Team}. R: A Language and Environment for Statistical Computing.
\newblock Vienna, Austria; 2018.
\newblock Available from: \url{https://www.R-project.org}.

\bibitem{plspm2017gaston}
Sanchez G, Trinchera L, Russolillo G. plspm: Tools for Partial Least Squares
  Path Modeling (PLS-PM); 2017.
\newblock R package version 0.4.9.
\newblock Available from: \url{https://CRAN.R-project.org/package=plspm}.

\end{thebibliography}
\bibliographystyle{vancouver}

\end{document}